\documentclass[a4paper]{panle}
\usepackage{url}
\usepackage{cite}
\usepackage{wrapfig}
\usepackage{graphicx}
\usepackage{amssymb}
\usepackage{amsfonts}
\usepackage{amsmath}
\usepackage{longtable}
\usepackage{rotating}
\usepackage{lscape}
\usepackage{epsfig}
\usepackage{mathrsfs}
\usepackage{multirow}
\usepackage{multicol}
\usepackage{bm}
\usepackage{mathptmx}

\usepackage{lineno}

\originalTeX
\begin{document}

\title{Analytic continuation of the $T$- and $S$-matrices\\ to unphysical sheets in a multichannel extension\\ of the Friedrichs-Faddeev model}
\maketitle
\authors{M.\,V.~Kremenetsky\,$^{a,b,}$\footnote{E-mail: kremisha2004@gmail.com}, A.\,K.~Motovilov$\,^{a,b,}$\footnote{E-mail: motovilv@theor.jinr.ru}}
\setcounter{footnote}{0}
\from{$^{a}$\,Dubna State University, 19 Universitetskaya, 141980 Dubna, Russia}
\from{$^b$\,Bogolyubov Laboratory of Theoretical Physics, JINR, 6 Joliot-Curie, 141980 Dubna, Russia}


\begin{abstract}
We consider a matrix multichannel extension of the Friedrichs-Faddeev model. First, we formulate conditions on the kernels of the potential
matrix that guarantee for the $T$-matrix (and then for $S$-matrix) the existence
of analytical continuation to unphysical energy sheets. We also obtain explicit
representations for the analytically continued $T$- and $S$-matrices on unphysical
sheets in terms of these matrices themselves, but only taken in the physical sheet.
From the explicit representations obtained, it follows that the resonances on a
certain unphysical sheet are nothing but the points at which the corresponding truncation of the scattering matrix in the physical sheet has an
eigenvalue of 0.
\bigskip

\noindent\textit{\small To be published in Physics of Particles and Nuclei (PEPAN),\\ https://www.pleiades.online/en/journal/physpart/.}

\end{abstract}
\vspace*{6pt}

\label{sec:intro}
\section*{INTRODUCTION}

The Friedrichs-Faddeev model \cite{Faddeev:1964}, which is an
extension of the Friedrichs model \cite{Friedrichs:1948},
corresponds to the Hamiltonian $h=h_0+V$ where $h_0$ stands for the
operator of multiplication by the independent variable in the
Hilbert space $\mathscr H=\ L_2([a,b],\mathfrak h)$ with $\mathfrak
h$ an auxiliary Hilbert space and $-\infty\leq a<b\leq \infty$:
\begin{equation}
    (h_0f)(\lambda)=\lambda f(\lambda),\quad f\in \mathscr H,\quad\lambda\in[a,b].
\end{equation}
The perturbation $V$ is an integral operator in $\mathscr H$:
\begin{equation}
    (Vf)(\lambda)=\int_a^{b}  v(\lambda,\mu)f(\mu)\,d\mu,\quad f\in \mathscr H,
\end{equation}
where the value of $v(\lambda,\mu)$ is assumed to be a compact operator in
$\mathfrak h$ and $v(\lambda,\mu)=v(\mu,\lambda)^*$, with a reasonably
good dependence on the variables $\lambda,\mu\in [a,b]$.

In the present work, the entries $h_0$, $V$ and $h$ of the
Friedrichs-Faddeev model are replaced with the respective block
operator matrices
    \begin{equation}
    \bm{H_0}=\left(\begin{matrix}
    h_1&0&\dots&0\\
    0&h_2&\dots&0\\
    \vdots&\vdots&\ddots&\vdots\\
    0&0&\dots&h_n
\end{matrix}\right)\text{, \qquad}\bm{V}=\left(\begin{matrix}
    v_{1,1}&v_{1,2}&\dots&v_{1,n}\\
    v_{2,1}&v_{2,2}&\dots&v_{2,n}\\
    \vdots&\vdots&\ddots&\vdots\\
    v_{n,1}&v_{n,2}&\dots&v_{n,n}
\end{matrix}\right),
\end{equation}
   \begin{equation}
        \bm{H}=\bm{H}_{0}+\bm{V},
    \end{equation}
considered as operators in the Hilbert space
$\displaystyle \mathscr H:=\bigoplus_{i=1}^n \mathscr H_i$ with
$\mathscr H_i= L_2([a_i,b_i],\mathfrak h_i)$
where $\mathfrak h_i$, $i=1,2,\ldots,n$, are auxiliary Hilbert
spaces (it is assumed that the intervals $[a_i,b_i]$ are all
different, i.e. $[a_i,b_i]\neq [a_j,b_j]$ whenever $i\neq j$). The
entries $h_i: \mathscr H_i\rightarrow \mathscr H_i$ and $v_{i,j}:
\mathscr H_j\rightarrow \mathscr H_i$, $i,j=1$, \dots, $n$, are defined as
follows
\begin{equation}
(h_if_i)(\lambda_i)=\lambda_if_i(\lambda_i),\text{\hspace{0.5cm}}(v_{i,j}f_j)(\lambda_i)=
\int_{a_j}^{b_j}v_{ij}(\lambda_i,\nu_j)f_j(\nu_j)\,d\nu_j.
\end{equation}
Here $\lambda_i\in[a_i,b_i]$, $\nu_j\in[a_j,b_j]$, $f_j\in
L_2([a_j,b_j],\mathfrak h_j)$, and the value of
$v_{ij}(\lambda_i,\nu_j)$ is assumed to be a compact operator from
$\mathfrak{h}_j$ to $\mathfrak{h_i}$. We also assume, just for
simplicity of the presentation, that all the functions
$v_{ij}(\lambda,\mu)$ are defined and analytic (with respect to the
$\mathcal{B}(\mathfrak{h}_j,\mathfrak{h}_i)$-topology) in both
$\lambda$ and $\mu$ in some domain $\Omega\subset\mathbb{C}$
containing simultaneously all the intervals $[a_i,b_i]$,
$i=1,2,\ldots,n$. Furthermore, it is supposed that
$v_{ij}(\lambda,\mu)=v_{ji}(\mu,\lambda)^*$, whenever
$\lambda,\mu\in\Omega\cap\mathbb{R}$, thus, making the operator $\bm
V$ (and hence the operator $\bm H$) self-adjoint.

In this work, we derive explicit representations for the
analytically continued ${T}$- and ${S}$-matrices to unphysical
energy sheets in terms of these matrices themselves, but only taken
on the physical sheet, thus extending the result of
\cite{Motovilov:2019} previously obtained for the original
Faddeev-Friedrichs model. Similarly to \cite{Motovilov:2019} and
\cite{Motovilov}, we  show that the resonances on a certain
unphysical sheet are the points on the physical sheet where the
correspondingly truncated ${S}$-matrix has an
eigenvalue of zero.

Throughout the work by $\mathcal{B}(\mathfrak{h},\mathfrak{g})$ we
denote the Banach space of bounded operators from a Hilbert space
$\mathfrak{h}$ to a Hilbert space $\mathfrak{g}$.

\section*{ANALYTIC CONTINUATIONS OF THE $T$- AND $S$-MATRICES}


Denote by $\bm{R}_0$ and $\bm R$ the resolvents of the operators $\bm{H}_0$ and $\bm H$, respectively:    ${\bm{R}_0}(z)=({\bm{H}_0}-z\bm{I})^{-1}$, $z\in\rho(\bm{H}_0)$, ${\bm{R}}(z)=({\bm{H}}-z\bm{I})^{-1}$,
$z\in\rho(\bm{H})$, where $\rho(A)$ stands for the resolvent set of
an operator $A$. Recall that the $T$-matrix is introduced in the
following way:
 \begin{equation}\label{T}
        {\bm T}(z):={\bm V}-{\bm V}{\bm R}(z){\bm V},\quad  z\in\rho(\bm H).
    \end{equation}
It satisfies the two variants of the corresponding
Lippmann-Schwinger equation:
    \begin{equation}\label{ls1}
{\bm{T}({z})=\bm{V}-\bm{V}\bm{R}_0({z})T({z})}, \quad {\bm{T}({z})=\bm{V}-\bm{T}({z})\bm{R}_0({z})\bm{V}}, \quad z\in \rho(\bm{H}_0).
\end{equation}
Usually, it is convenient to rewrite the equations in \eqref{ls1} in
terms of kernels:
\begin{align}
\label{t1}
    t_{ij}(\lambda_i,\mu_j,z)&=v_{ij}(\lambda_i,\mu_j)-\sum_{k=1}^n\int_{a_k}^{b_k}  v_{ik}(\lambda_i,\nu_k)\dfrac{1}{\nu_k-z}t_{kj}(\nu_k,\mu_j,z)\, d\nu_k,\\
\label{t2}
    t_{ij}(\lambda_i,\mu_j,z)&=v_{ij}(\lambda_i,\mu_j)-\sum_{k=1}^n\int_{a_k}^{b_k} t_{ik}(\lambda_i,\nu_k,z) \dfrac{1}{\nu_k-z}v_{kj}(\nu_k, \mu_j)\, d\nu_k.
\end{align}
Almost literally following the approach of \cite{Faddeev:1964}, one
may plunge these equations into an appropriate Banach space in such
a way that the Lippmann-Schwinger operators $\bm{V}\bm{R}_0({z})$
and $\bm{R}_0({z})\bm{V}$ turn out to be compact in this space. In
the case under consideration the role of such a Banach space is
played by a product space formed of $n$ channel H\"older classes
with H\"older continuity index $\alpha\in(1/2,1]$. This allows one
to study the analytic properties of the kernels
$t_{ij}(\lambda,\mu,z)$ with respect to all the involved variables
$\lambda,\mu$ and $z$.

Two things here are worth mentioning: Firstly, from \eqref{t1} one
concludes that since $v_{ij}(\lambda_i,\mu_j)$, $i,j=1$, \dots, $n$, are
analytic in $\lambda_i\in\Omega$, the same holds for
$t_{ij}(\lambda_i,\mu_j,z)$.  From \eqref{t2}, a similar conclusion
should be done with respect to the holomorphy of the functions
$t_{ij}(\lambda_i,\mu_j,z)$ in respective variables
$\mu_j\in\Omega$. Secondly, the equations \eqref{t1}, \eqref{t2}
include Cauchy type integrals, which can be easily continued through
the cuts along the respective segments $[a_k,b_k]$:
    \begin{equation}\label{c1}
        F(z)=\int_\alpha^\beta \frac{f(x)}{x-z}\,dx \longrightarrow
        F'_l(z)=F(z)-2\pi i lf(z),\quad l\in\mathbb Z,
    \end{equation}
provided that $f$ is holomorphic in an open neighborhood of $[\alpha,\beta]$.
The number $l$ depends on the direction of the continuation
path and counts the crossings of the segment $[\alpha,\beta]$ during the
rounds of $z$ about the end points $\alpha$ and/or $\beta$ (see, e.g.,
\cite{Motovilov:2019,Motovilov}).

In \eqref{t1} and \eqref{t2}, there are Cauchy type integrals
associated with the intervals $[a_k,b_k]$, and every end point $a_k$
and $b_k$, $k=1,2,\ldots,n$, turns into a logarithmic branching point
of the Riemann surface. Thus, to identify an unphysical energy sheet
we introduce the multi-index $L=(l_1,l_2,\dots,l_n)$, where $l_k\in
\mathbb Z$ corresponds to residues resulting from continuations of
the participating integrals through $[a_k,b_k]$, $k=1$, \dots, $n$. As a
result, we get from \eqref{t1} the following equations for
analytically continued $t_{ij}(\lambda_i,\mu_j,z)$ (similar results
could be obtained for \eqref{t2}, but from now on we restrict
ourselves to the results for \eqref{t1}):
\begin{equation}\label{15}
\begin{array}{c}
\displaystyle t^L_{ij}(\lambda_i,\mu_j,z)=v_{ij}(\lambda_i,\mu_j)-\\
\displaystyle -\sum_{k=1}^n\int_{a_k}^{b_k}  v_{ik}(\lambda_i,\nu_k)\dfrac{1}{\nu_k-z}(\nu_k,z)t^L_{kj}(\nu_k,\mu_j,z) \,d\nu_k+\\
\displaystyle+ \sum_{k=1}^n l_{k} 2\pi i
v_{ik}(\lambda_i,z)t^L_{kj}(z,\mu_j,z).
\end{array}
\end{equation}

In order to rewrite \eqref{15} in operator form, we introduce the
operators $J=(j_1,j_2,\dots, j_n)$ and $J^+=(j_1^+,j_2^+,\dots, j_n^+)$,
which act as follows:
\begin{equation}
(j_iA_{ik}(\cdot,\mu))(z):=A_{ik}(z,\mu), \quad
(A_{ik}(\lambda,\cdot)j_i^+)(z):=A_{ik}(\lambda,z),
\end{equation}
that is, perform setting of the corresponding argument of an
analytic two-argument
$\mathcal{B}(\mathfrak{h}_k,\mathfrak{h}_i)$-valued function
$A_{ik}(\lambda,\mu)$ on the energy shell.

Equation \eqref{15} in terms of operators in $\mathscr H$ and
operators $J$, $J^+$ will be:
\begin{equation}
{\bm T}_L(z)={\bm V}-{\bm V}{\bm R}_0(z){\bm T}_L(z)+2\pi i {\bm
V}J^+{\bm L}J{\bm T}_L(z).
\end{equation}
Here, ${\bm L}=\text{diag}(l_1,l_2,\dots,l_n)$, and ${\bm T}_L(z)$
denotes the continuation of the ${T}$-matrix. By introducing the
projection ${\hat{\bm{L}}}=\bigl(\hat{L}_{ij}\bigr)_{i,j=1}^n$ with
$\hat{L}_{ij}=0$ if ${L}_{ij}=0$ and $\hat{L}_{ij}=1$ if
${L}_{ij}\neq0$ we, after some transformations, obtain
\begin{equation}
\bm{T}_{L}({z})=\bm{T}({z})+2\pi {i} \bm{T}({z}){J}^+ \bm{L}\bm{S}_{L}^{-1}({z})\hat{\bm{L}}{J}\bm{T}({z}).
\end{equation}
Here  ${\bm S}_L(z)={\bm I}-2\pi i\hat{{\bm L}}J{\bm T}(z)J^+{\bm
L}$ is a truncated scattering matrix and it is taken in the physical
sheet taken. We have thus obtained a representation of the analytic
continuation of the $T$-matrix, as well as showed that the
resonances of the $\bm{T}_{L}$ are the points where $\bm{S}_{L}$ has an eigenvalue of zero.

We can also continue $\bm{S}_{L}$ to an unphysical sheet
$\Pi_{L'}$:
\begin{equation}
\label{SS} \left.{\bm S}_{L}(z)\right|_{\Pi_{{L'}}}={\bm I}-2\pi i
\hat{{\bm L}} J{\bm T}_{{L'}}(z)J^+{\bm L} ={\bm I} -{\bm
S}_{{L'}}^{-1}(z)({\bm S}_L-{\bm I}),
\end{equation}
with a notable special case $L'=L$ where $\left.{\bm
S}_{L}(z)\right|_{\Pi_{{L}}}= {\bm S}_L^{-1}(z)$.

From \eqref{SS} it follows that the poles of $S_L(z)|_{\Pi_L'}$
are found at the points $z$ where $S_{L'}(z)$ (considered in the
physical energy sheet) have an eigenvalue of zero.

\section*{CONCLUSIONS}
We have derived the explicit representations for the analytically
continued ${T}$- and ${S}$-matrices in terms of these matrices
themselves, but taken on the physical sheet. We also showed that the
resonances on a certain unphysical sheet are nothing but the points
where the corresponding truncated ${S}$-matrix on the physical sheet
has an eigenvalue of zero.

\section*{CONFLICT OF INTEREST}
The authors declare that they have no conflicts of interest.

\bibliographystyle{pepan}
\bibliography{K04_biblio}

\end{document}